\documentclass[12pt]{article}
\usepackage{epsf}
\usepackage[cp1251]{inputenc}
\usepackage[english]{babel}

\begin{document}

\title{
{\bf Current stage of understanding and description of hadronic elastic diffraction}}
\author{A.A. Godizov\thanks{E-mail: anton.godizov@gmail.com}\\
{\small {\it Institute for High Energy Physics, 142281 Protvino, Russia}}}
\date{}
\maketitle

\vskip-1.0cm

\begin{abstract}
Current stage of development of the high-energy elastic diffractive scattering phenomenology is reviewed. Verification of various theoretical models via comparison of 
their predictions with the recent D0 and TOTEM data on the nucleon-nucleon total and differential cross-sections is presented.
\end{abstract}

\section*{Introduction}

Diffractive phenomena in hadron physics are related to strong interaction. QCD is recognized as the fundamental theory of strong interaction. Thus, description of 
diffractive reactions at high energies should be grounded on some QCD-based techniques. But the special status of diffractive studies at high-energy colliders is determined 
by the fact that diffraction of hadrons takes place due to interaction at large distances. Indeed, the transverse size of the hadron interaction region can be estimated 
through the corresponding Heisenberg uncertainty relation, and extraction of this quantity from experimental elastic angular distributions can be done without dealing with 
any theory. For example, at the SPS, Tevatron, and LHC energies it is of order 1 fm. Technically, this means that we are in the so-called ``non-perturbative regime'' and 
straight applying QCD to description of hadronic diffraction is disabled, since QCD, at its current stage of development, has no essential progress outside of 
perturbative calculations.

Hence, one is enforced to invent ``plausible'' models which bear, at least, general QCD properties, as much as possible.

\section*{Scattering amplitude, Born term (``eikonal'') and Regge trajectories}

In the vast majority of diffraction models the notion of reggeons (analytic continuations of resonance spectra) is used.

Below, the recipe for calculation of the elastic scattering amplitude in the framework of the Regge-eikonal approach \cite{collins} is presented: 
$$
T_{12\to 12}(s,t) = 4\pi s\int_0^{\infty}db^2J_0(b\sqrt{-t})\frac{e^{2i\delta_{12\to 12}(s,b)}-1}{2i}\,,
$$
\newpage
$$
\delta_{12\to 12}(s,b) = \frac{1}{16\pi s}\int_0^{\infty}d(-t)J_0(b\sqrt{-t})\delta_{12\to 12}(s,t) = \frac{1}{16\pi s}\int_0^{\infty}d(-t)J_0(b\sqrt{-t})\times
$$
$$
\times\left\{\sum_n\left(i+{\rm tg}\frac{\pi(\alpha_n^+(t)-1)}{2}\right)\Gamma_n^{(1)+}(t)\Gamma_n^{(2)+}(t)s^{\alpha_n^+(t)}\mp\right.
$$
$$
\left.\mp\sum_n\left(i-{\rm ctg}\frac{\pi(\alpha_n^-(t)-1)}{2}\right)\Gamma_n^{(1)-}(t)\Gamma_n^{(2)-}(t)s^{\alpha_n^-(t)}\right\}\,.
$$
Here $s$ and $t$ are the Mandelstam variables, $b$ is the impact parameter, $T$ is the elastic scattering amplitude, eikonal $\delta$ is the sum of single-reggeon-exchange 
terms, $\alpha(t)$ are Regge trajectories, and $\Gamma(t)$ are the reggeonic form-factors of colliding particles. Besides the Regge-eikonal approximation, there exist 
various approaches exploiting the notion of reggeons.

Calculation of Regge trajectories within QCD is one of the main theoretical problems of hadron physics. What are fundamental achievements in this direction?

\section*{Regge trajectories and QCD}

The first approach yielding some results on QCD Regge trajectories is the famous BFKL approach based on solving the so-called BFKL equation \cite{kuraev} which is some 
modification of the Bethe-Salpeter equation. Within this approach, the behavior of the Regge trajectories of the reggeons composed of two reggeized partons was calculated 
at asymptotically high transfers (in the case of quark-antiquark pair by Kwiecinski \cite{kwiecinski} and for two gluons by Kirschner and Lipatov \cite{lipa}):
$$
\alpha_{\bar qq}(t) = \sqrt{\frac{8}{3\pi}\alpha_s(\sqrt{-t})}+o(\alpha_s^{1/2}(\sqrt{-t}))\,,
$$
$$
\alpha_{gg}(t) = 1+\frac{12\,\ln 2}{\pi}\alpha_s(\sqrt{-t})+o(\alpha_s(\sqrt{-t}))\,,
$$
where $\alpha_s$ is the QCD running coupling.

As well, the intercept of the leading Regge trajectory, pomeron, was calculated by Fadin and Lipatov and, also, Ciafaloni and Camici \cite{camici}: 
$$
\alpha_{gg}(0) = 1+\frac{12\,\ln 2}{\pi}\alpha_s(\mu)\left(1-\frac{20}{\pi}\alpha_s(\mu)\right)+o(\alpha^2_s(\mu))\,.
$$
In contrast to the asymptotic relations at high transfers, the last expression is explicitly not renorm-invariant, since it depends on an arbitrary renormalization scale 
$\mu$. Certainly, though Regge trajectories as analytic (i.e. unique) continuations of resonance spectra should be renorm-invariant, they could be approximated by some 
renorm-noninvariant expressions (like in QED), but the proper choice of the renormalization scheme and scale should be physically motivated. Also, the theoretical 
uncertainty (determined by the neglected terms) should be low enough for a possibility of practical use of the obtained renorm-noninvariant approximations under 
calculation of scattering amplitudes. At the current stage of the BFKL approach development, the theoretical uncertainty of the BFKL pomeron intercept value is rather 
high, and, thus, the BFKL method provides information only about the asymptotic behavior of QCD Regge trajectories at ultra-high values of transferred momentum.

The second approach is the less-known Lovelace approach \cite{lovelace} which deals with the Bethe-Salpeter equation in some asymptotic regime, where we do not need 
reggeization of the partons composing the considered bound state. The main feature of this approach is exploitation of renorm-invariant kernels in the BS equation. As a 
result, we obtain renorm-invariant numbers for the intercepts of Regge trajectories. In his paper \cite{lovelace} Lovelace considered asymptotically free $\phi^3_6$-theory 
and found some infinite series of intercepts:
$$
(\alpha^{(k)}_{\phi\phi}(0)+1)(\alpha^{(k)}_{\phi\phi}(0)+2)(\alpha^{(k)}_{\phi\phi}(0)+3)=\frac{16}{3(2k+1)}
$$
with $\alpha^{(0)}_{\phi\phi}(0)\approx -0.06273$. An analogous result was obtained for some series of meson Regge trajectories in QCD \cite{godizov}:
$$
\alpha^{(k)}_{\bar q q}(0)=\frac{9(N_c^2-1)}{(2k+1)N_c(11N_c-2n_f)}-1
$$
(if $N_c=3$ and $n_f=6$, then $\alpha^{(0)}_{\bar q q}(0)=1/7$). Note, that in the Lovelace approach the intercepts of Regge 
trajectories do not depend on the coupling at all, what is a general consequence of renorm-invariance in massless field theories \cite{petrov}.

Unfortunately, due to technical difficulties, the leading gluon-gluon and quark-antiquark series (corresponding to the trajectories with the Kwiecinski and 
Kirschner-Lipatov asymptotical behavior) were not calculated, though the existence of some infinite series condensing to $\alpha^{(\infty)}_{g g}(0)=1$ from above was 
proved \cite{heckathorn}.

Thus, we have got information about quantitative behavior of the QCD leading Regge trajectories at very high momentum transfers only. This region of transferred momenta, 
however, gives a negligible contribution to diffractive cross-sections. In other words, at present moment, QCD does not provide any quantitative result which could be 
directly used under construction of phenomenological models of hadronic diffraction, though some models try to adopt qualitative QCD features. 

\section*{Phenomenological models}

The phenomenological models of the nucleon-nucleon elastic diffractive scattering, proposed before the TOTEM preliminary results had been published, could be divided into 
2 groups: the models exploiting the notion of reggeons \cite{jss, dgj,dgm,pp,bsw,m,agn,mn,jll,glm,rmk,o,g} and the non-reggeon models \cite{acmm,ilp,fgl,bd,bh}. 

In several reggeon models the eikonal representation of the scattering amplitude is used, where the eikonal is the sum of single-reggeon exchange terms with a supercritical 
pomeron (or pomerons) as the leading term \cite{pp,bsw,g}. Other reggeon models do not use the eikonal representation but introduce some complicated leading Regge 
singularities: the pomeron as double Regge pole in \cite{jss, dgj,dgm,m,jll}, and the so-called froissaron, the leading Regge cut, in \cite{agn,mn}. Also, these models 
take account of the contributions from secondary Regge poles.

There exists a separate subgroup of reggeon models \cite{glm,rmk,o} exploiting the methods of the Reggeon Field Theory. The eikonal here is replaced by the so-called 
opacity which is the sum of not only single-reggeon-exchange terms but, also, multi-reggeon exchanges. Low-mass dissociation in the intermediate states is taken into 
account as well.

The non-reggeon phenomenological schemes could be divided into the models not appealing to QCD \cite{acmm,bd} and the so-called QCD-inspired models \cite{ilp,fgl,bh}. Model 
\cite{acmm} exploits the general principles and the derivative dispersion relations as extra conditions. Model \cite{bd} is some variant of the quasi-potential approach. In 
\cite{ilp} nucleon is considered to have an outer cloud of quark-antiquark pairs, an inner shell of baryonic charge, and a central quark bag containing the valent 
quarks (small-angle scattering is due to overlapping of the outer clouds and the exchange by $\omega$-meson). In the Dipole Cascade Model \cite{fgl} nucleon is introduced 
as color dipole and interaction between hadrons at ultra high energies is presumed to be dominated by perturbative effects. Model \cite{bh} uses the eikonal composed of 3 
terms which are called the contributions from the quark-quark, quark-gluon, and gluon-gluon interaction, though the corresponding expressions are not derived from QCD 
directly and contain numerous free parameters.

All the mentioned models give different predictions for the nucleon-nucleon total and differential cross-sections at the LHC.

\section*{Models vs. D0 and TOTEM}

Comparison of the model predictions for the $pp$ total cross-section with the value measured by the TOTEM Collaboration \cite{totemtot} reveals that some of the models may 
be judged as discriminated already at this stage (Tab. \ref{totem}). Though others survive.
\begin{table}[ht]
\begin{center}
\begin{tabular}{|l|l|}
\hline
{\bf The Model}   &  $\sigma^{pp}_{tot} (7\,TeV)$, mb \\
\hline
P. Desgrolard, M. Giffon, L.L. Jenkovszky,  & 87 ({\it 6 TeV}) \\
$\;\;\;$ Z. Phys. C {\bf 55} (1992) 637   & \\
\hline
A. Donnachie, P.V. Landshoff, Phys. Lett. B {\bf 296} (1992) 227  & 91  \\
\hline
P. Desgrolard, M. Giffon, E. Martynov,  & 95  \\
$\;\;\;$ Eur. Phys. J. C {\bf 18} (2000) 359   & \\
\hline
V.A. Petrov, A.V. Prokudin, Eur. Phys. J. C {\bf 23} (2002) 135   & 97 $\pm$ 4  \\
\hline
C. Bourrely, J. Soffer, T.T. Wu, Eur. Phys. J. C {\bf 28} (2003) 97   & 93  \\
\hline
R.F. Avila, S.D. Campos, M.J. Menon, J. Montanha,   & 94 \\
$\;\;\;$ Eur. Phys. J. C {\bf 47} (2006) 171   &   \\
\hline
M.M. Islam, R.J. Luddy, A.V. Prokudin,  & 97.5\\
$\;\;\;$ Int. J. Mod. Phys. A {\bf 21} (2006) 1  & \\
\hline
E. Martynov, Phys. Rev. D {\bf 76} (2007) 074030  & 91 \\
\hline
R.F. Avila, P. Gauron, B. Nicolescu, Eur. Phys. J. C {\bf 49} (2007) 581  & 108 \\
\hline
E. Martynov, B. Nicolescu, Eur. Phys. J. C {\bf 56} (2008) 57  & 95 \\
\hline
C. Flensburg, G. Gustafson, L. L\"onnblad,  & 98 $\pm$ 9 \\
$\;\;\;$ Eur. Phys. J. C {\bf 60} (2009) 233 &  \\
\hline
P. Brogueira, J. Dias de Deus, J. Phys. G {\bf 37} (2010) 075006  & 110 \\
\hline
M.M. Block, F. Halzen, Phys. Rev. D {\bf 83} (2011) 077901  & 95.5 $\pm$ 1 \\
\hline
L.L. Jenkovszky, A.I. Lengyel, D.I. Lontkovskyi,  & 98 $\pm$ 1 \\
$\;\;\;$Int. J. Mod. Phys. A {\bf 26} (2011) 4755 &   \\
\hline
E. Gotsman, E. Levin, U. Maor, Eur. Phys. J. C {\bf 71} (2011) 1553  & 91  \\
\hline
M.G. Ryskin, A.D. Martin, V.A. Khoze,   & 89  \\
$\;\;\;$ Eur. Phys. J. C {\bf 71} (2011) 1617 &  \\
\hline
S. Ostapchenko, Phys. Rev. D {\bf 83} (2011) 014018  & 93 \\
\hline
A. Godizov, Phys. Lett. B {\bf 703} (2011) 331  & 110  \\
\hline
\hline
\hline
The TOTEM Collaboration, Europhys. Lett. {\bf 101} (2013) 21002  & $98.0\pm 2.5$  \\
\hline
\end{tabular}
\end{center}
\caption{Comparison of the model predictions for the $pp$ total cross-section at $\sqrt{s} = $ 7 TeV with the TOTEM result.}
\label{totem}
\end{table}

However, some theorists may consider deviations in several percents from experimental data to be not fatal for any model. In any case, measurement of total cross-section 
only is not enough for proper discrimination among the models. 

\begin{figure}[ht]
\epsfxsize=8cm\epsfysize=7cm\epsffile{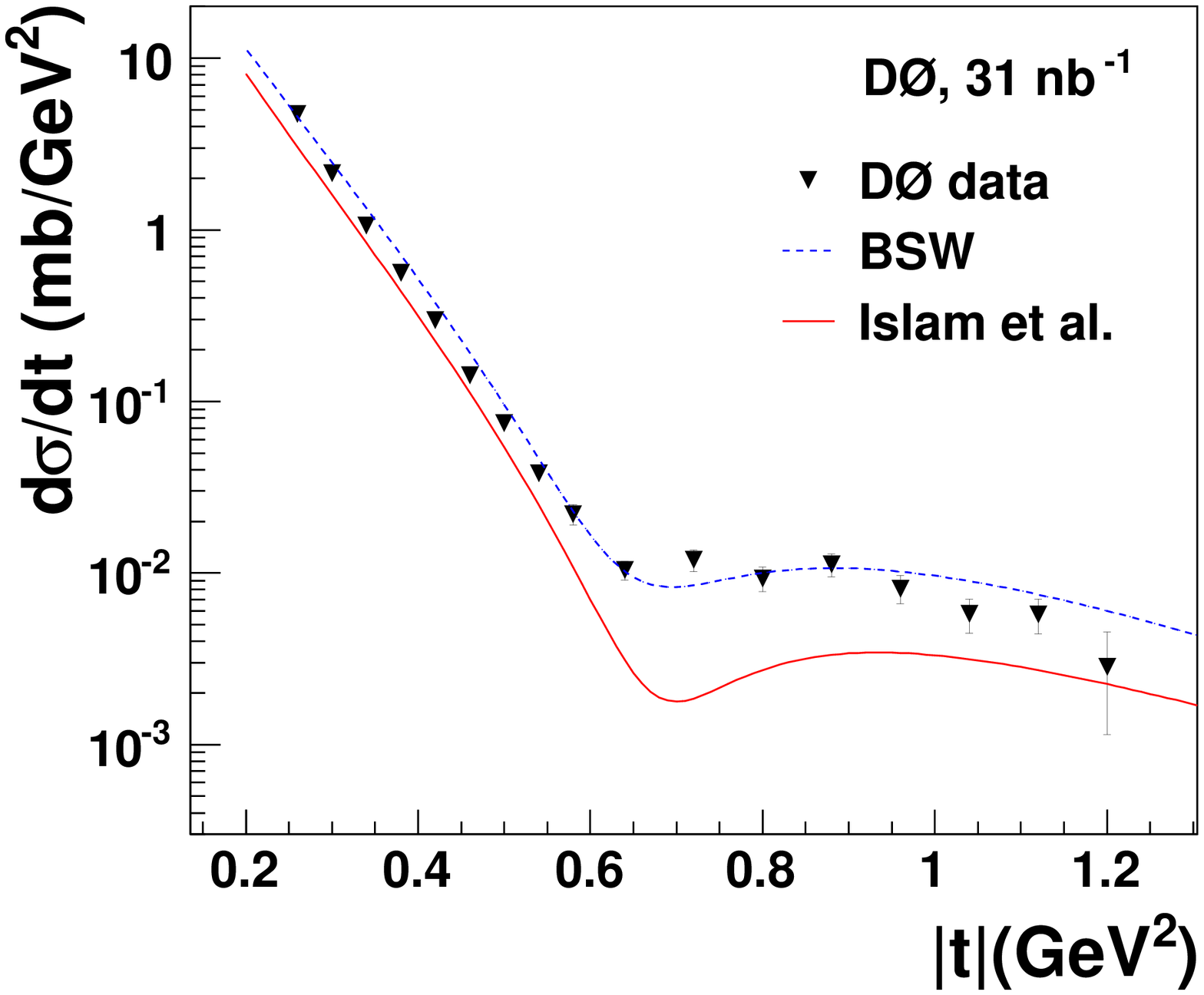}
\vskip -7.1cm
\hskip 9cm
\epsfxsize=7cm\epsfysize=7cm\epsffile{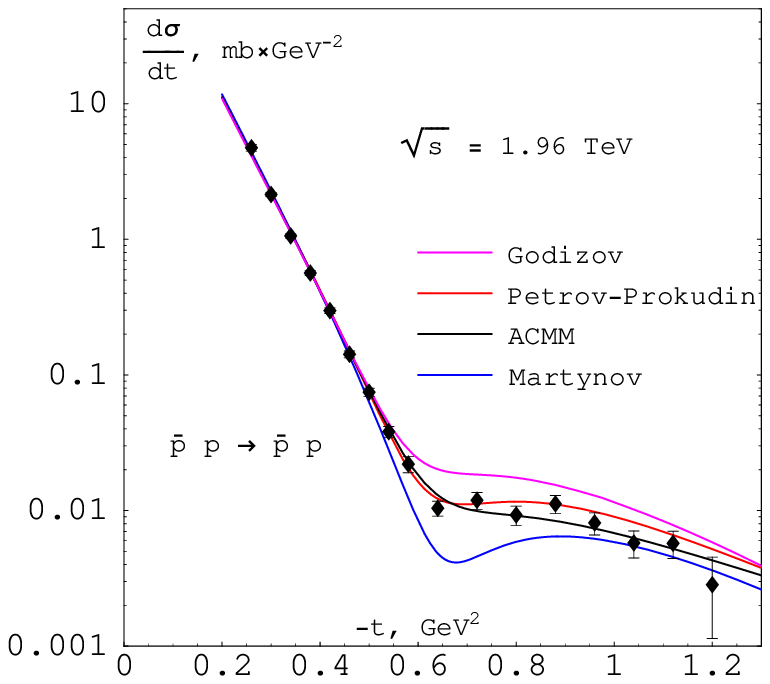}
\caption{Comparison of the model predictions for the $\bar pp$ differential cross-section at $\sqrt{s} = $ 1.96 TeV with the D0 results (the left picture is taken from 
\cite{d0}).}
\label{d0imp}
\end{figure}

Some models can be discriminated via comparison with the recently published D0 data on the $\bar pp$ differential cross-section \cite{d0} (Fig. \ref{d0imp}).

\begin{figure}[ht]
\epsfxsize=7.4cm\epsfysize=7.4cm\epsffile{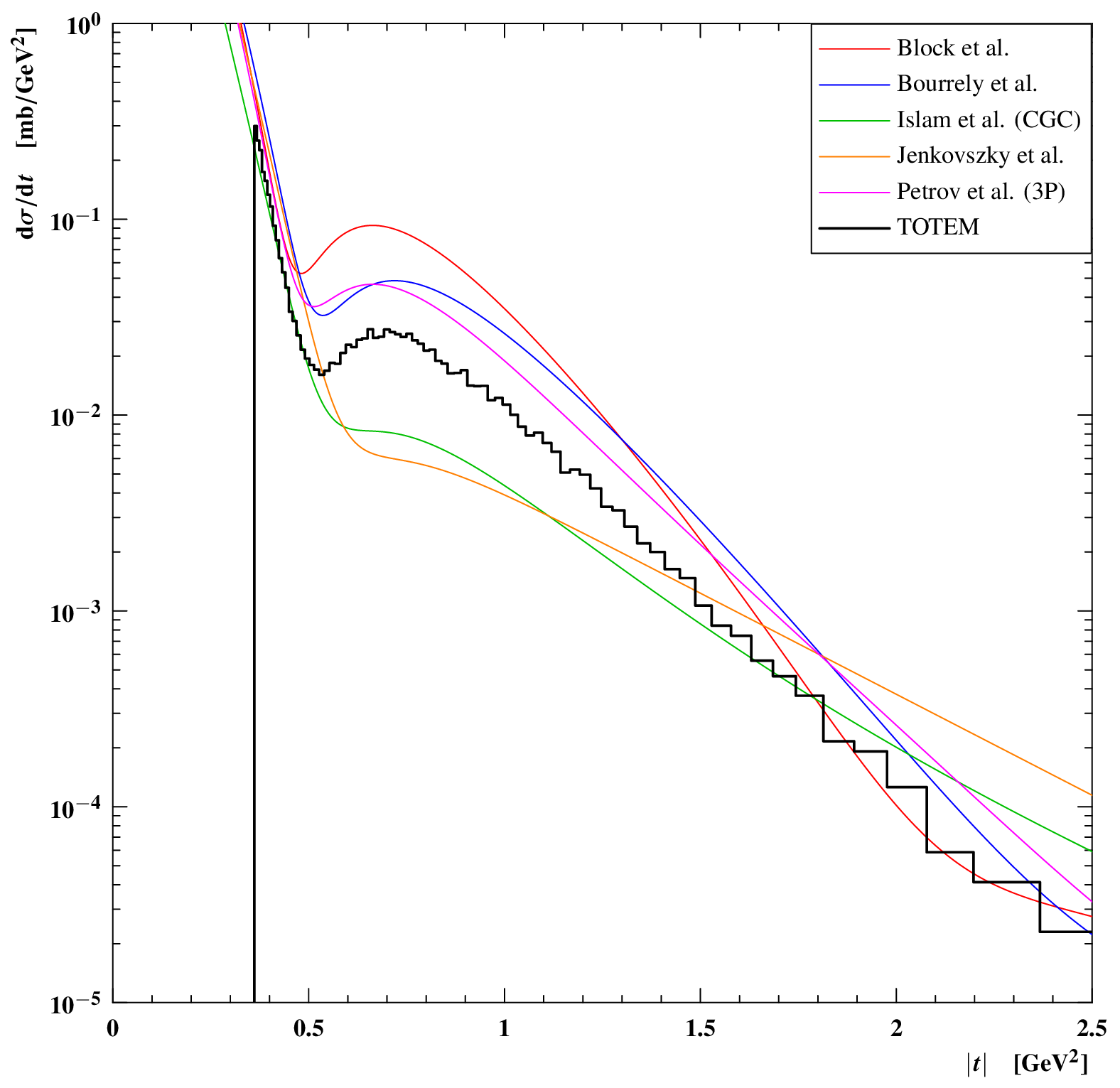}
\vskip -7.8cm
\hskip 8.5cm
\epsfxsize=8cm\epsfysize=8cm\epsffile{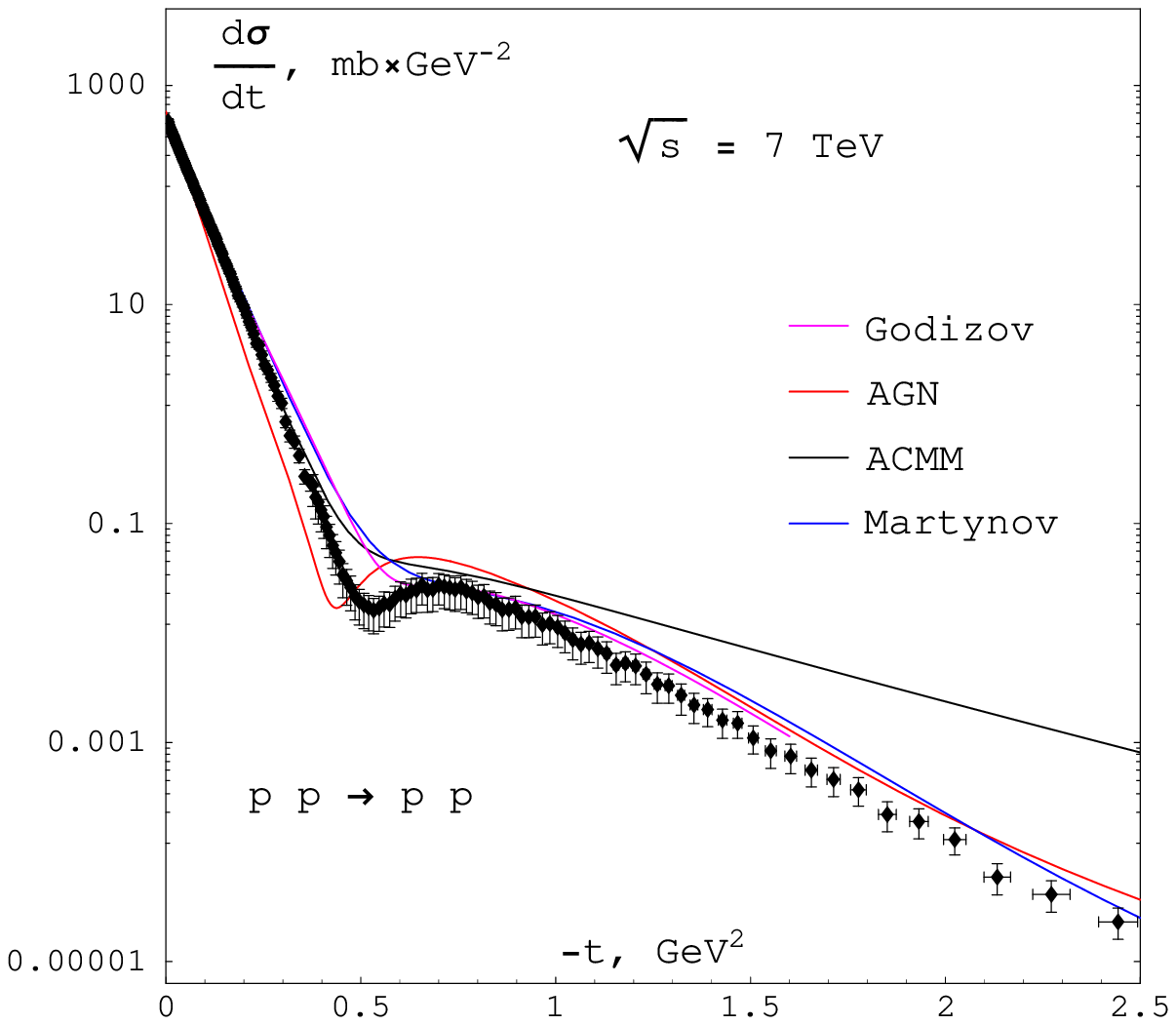}
\caption{Comparison of the model predictions for the $pp$ differential cross-section at $\sqrt{s} = $ 7 TeV with the TOTEM results (the left picture is taken from 
\cite{totemdiff}).}
\label{totimp}
\end{figure}

In their turn, the TOTEM data on the $pp$ differential cross-section \cite{totemtot,totemdiff} reveal a very strong discriminative power (Fig. \ref{totimp}). Inversely, the 
predictive efficiency of all of the considered models turned out very weak. I mean the fact that many models give a nice description of the nucleon-nucleon differential 
cross-sections from the ISR to the Tevatron energies \cite{diffexp} (with the collision energy increase in several tens of times). But though the ratio of the LHC energy to 
the Tevatron energy is only about 4, we can observe a huge discrepancy between the model curves and the data (for some models --- tens of percent, for others --- several 
times). 

Another question: what are the consequences of such a result for QCD? Is it falsified? Certainly, no. Only the phenomenological models are the subjects of falsification but 
QCD is not. Such a situation takes place due to the fact that all the hadron diffraction models are not grounded on analytical derivations from QCD, though some of them use 
adopted QCD terminology. 

In 2 years after the TOTEM preliminary results \cite{totemdiff} had been published we returned to the same stage as before the TOTEM measurements: there are numerous models 
\cite{w} -- \cite{dl} (very different by physical ground) which describe the TOTEM data more or less satisfactorily. But who can guarantee that such a simultaneous 
failure will not be reproduced after the forthcoming measurements at 14 TeV?

\section*{Conclusion}

We need a deeper interrelation of phenomenological models with QCD. First of all, we need development of some non-perturbative techniques for calculation of Regge 
trajectories in the diffractive (large distance) domain of QCD.

\end{document}